\documentclass[aps,twocolumn,amsmath,amssymb,reprint]{revtex4}

\usepackage{graphicx}
\usepackage{amsmath}
\usepackage{latexsym}
\usepackage{amsmath}
\usepackage{amssymb}
\usepackage{float}
\usepackage{color}
\usepackage{epstopdf}
\usepackage{hyperref}
\hypersetup{urlcolor=red,citecolor=blue}
\usepackage{bm}
\hypersetup{urlcolor=red,citecolor=blue}

\usepackage{epsfig}
\newcommand{\ztwo}{$\mathbb{Z}_2$ }

\begin{document}

\title{Robust electro-mechanical actuation in hydrogenated Xenes \\leading to reversible topological transition}

\author{Sujith Nedungattil Subrahmanian} 
\author{Nabendu Mondal}
\author{Joydeep Bhattacharjee}


\affiliation{School of Physical Sciences, National Institute of Science Education and Research, 
An OCC of Homi Bhabha National Institute, Odisha-752050,India}


\begin{abstract}
We report from first principles, the possibility of 
reversible onset of topological insulator(TI) phase in heavier hydrogenated Xenes (Xane), namely, 
germanane and stanane,
exclusively through in-plane electro-mechanical actuation.
It is found possible to systematically induce robust uniaxial strain
through non-uniform application of bias through gates of realizable length-scales
 embedded underneath.
%
Electrically induced  strain causes substantial lowering of band-gap across all Xanes,
eventually evolving through weak followed by strong topologically insulating phases 
beyond a threshold degree of bias in-homogeneity in heavier Xanes, 
promisingly within the range of bias sustained by the monolayers.
In case of nano-ribbons of these Xanes, bias applied in-homogeneously across width promises 
switchable emergence of TI phase over a fraction of width and 
topologically protected interface states localizable anywhere across the half-width of the ribbon.
The demonstrated electro-mechanical actuation and the associated topological tuning of band-structure,
thematically verified in gapped graphene based representative systems 
within the Kane-Mele model at half-filling,   
should be possible in the broader class of two dimensional covalent networks made of elements
of the $p$-block.
\end{abstract}
\maketitle
\section{Introduction}
\label{intro}
Efforts to evolve beyond the three dimensional silicon-metal combination for channel-drain$/$source assembly in active elements of electronic circuitry, began with graphene about a quarter of a century ago\cite{graphene2004,das2011electronic} with the promise of drastic reduction in dimensionality, dissipative loss and weight of devices. The efforts generalized into the broader family of two dimensional materials constituted by the elements of the $p$ block down the periodic table\cite{balendhran2015elemental}, namely the Xenes -
silicene\cite{tao2015silicene}, germanene\cite{acun2015germanene}, stanene\cite{lyu2019stanene} and further to the halcogenides\cite{Review_TMDCs,TMDC_devices}, and continues ever more vivaciously in recent years towards harnessing the exotic quantum phases\cite{review_HasanKane} in aid to ease of passage of charge and thermal management central to large scale integration of processors with fast increasing density. In this direction, two dimensional topological insulators(TI)\cite{review_atomthinQSH} has been in focus for more than a decade now, on account of the promise of  scatteringless conduction through the topologically protected edge-states  they host if cut into ribbons\cite{jin2023topological}. However, sustaining the topological protection in ambient working condition and configuration of devices is the key challenge to overcome, wherein modest progress has been reported \cite{kou2013graphene,han2017room,chen2024temperature} in recent years. In fact, evidence of possible tunability of topological edge states in germanene\cite{Germanene_QSH,klaassen2024tunability} at room temperature using electric field is a contemporary matter of interest from device perspective. TIs with reversible control over transition between the topological and trivial phases is thus envisaged to be of significant technological interests and the motivation of this work as well.

With opening of gap at Dirac point upon hydrogenation in both the sub-lattices, 
hydrogenated Xene monolayers (Xanes) -
graphane, silicane, germanane and stanane, offers direct band gaps at $\Gamma$,
with the optical gap in the visible range for germanane\cite{bianco2013stability,lu2017quasiparticle}.
%
Substituion of H by heavier atoms\cite{GeH_ChenSi,GeIH_Padilha} , 
or appropriate functional groups, combined with biaxial strain\cite{GeH_ChenSi,ma2014strain,GeCH3_QSH, GeCH3Si_ayan_dutta}, 
has been shown to nontrivially modulate band-gap of Xanes, 
leading to topological phase transition in the heavier ones, namely, germanane and stanane,
driven by the spin-orbit coupling(SOC) of primarily the $\sigma$ orbitals. 
Homogeneously strained Xanes are expected to undergo topological phase transition for the same reason,
since a stretched X-X bond would imply increase in enengy of the occupied $\sigma$ orbitals taking them 
closer to the edge of the valence band which may eventually facilitate to band inversion. 
On account of the increasing strength of SOC from Si to Sn, the degree of strain required for such transition progressively decreases from silicane to stanane.
As verified in this work for uniaxial strain, 
the degree of strain required for the transition 
in case of silicane is as high as above  25\%, reducing to almost half
of that and further to a quarter in case of germanane and stanane respectively.
However, in order to be able to 
modulate charge passage through matter or vary the degree of their interaction with circularly  polarized light
on account of existence of the topologically protected helical edge or interface states,
a reversible onset of TI  phase exclusively through revocable  physical functionalization is highly desirable. Accordingly, electrically induced onset of topological phase beyond silicane has been proposed for bilayers and multi-layers of Xanes \cite{GeH_efield_QSH,SnH_efield_QSH} driven by unequal lowering of energies of the 
$s$ and $p$ orbitals in electric field owing to their even and odd parities respectively, 
leading to a feasible means of achieving band-inversion through application 
of electric field uniformly perpendicular to the layers.

Harnessing electric field as a source of strain, in this work, we report from first principles, electro-mechanical actuation as a possible route for controllable onset of weak and strong TI phases in Xanes beyond silicane.
%
Substantial lowering of band-gap leading eventually to inversion of valence and conduction band edges at $\Gamma$,  
leading to onset of TI phase, is demonstrated computationally in germanane and stanane mono-layers 
as a result of in-homogeneous strain induced 
exclusively through application of in-homogenous bias potential through realizable gate configurations. 
%
%
Correspondingly demonstrated the possibility of emergence of chiral interface states in Xane ribbons with realistically applied bias.
This work thus brings forth the effectiveness of electro-mechanical actuation in nontrivial reversible tuning of electronic structure
facilitating a host of physical and chemical functionalities ranging from onset of exotic quantum phases to controlled cleavage of 
covalent bonds in monolayers of $p$ block elements through patterned  application of bias. 

\begin{figure}[t]
\flushleft (a)\vspace{-0.30cm}\\ \centering 
\hspace{-0.30cm}
\includegraphics[scale=0.38]{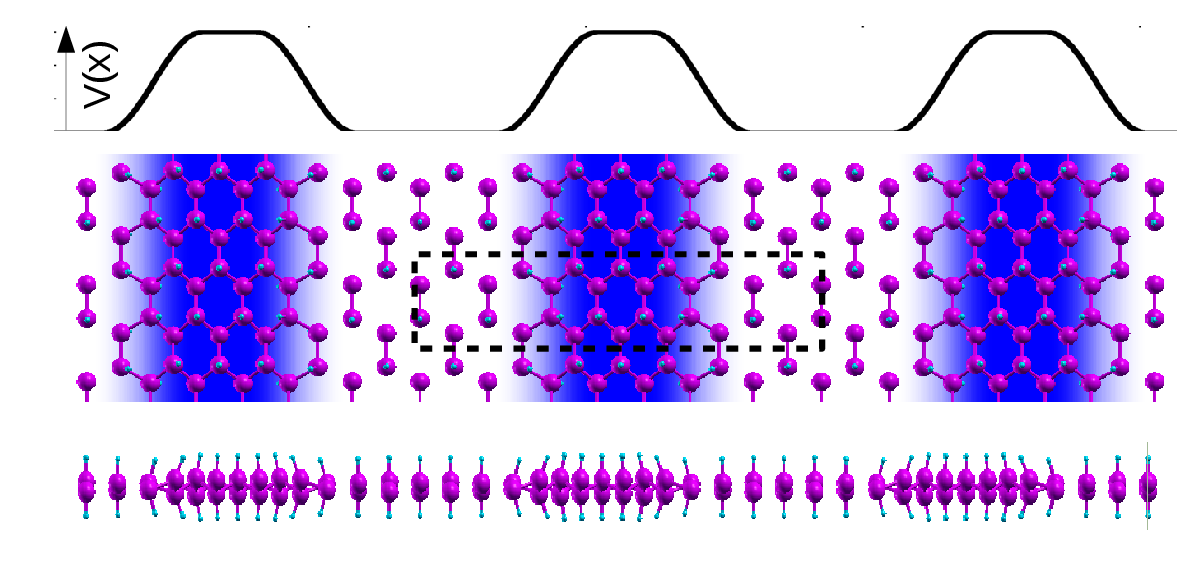}
\flushleft (b)\vspace{-0.40cm}\\ \centering 
\hspace{0.30cm}
\includegraphics[scale=0.41]{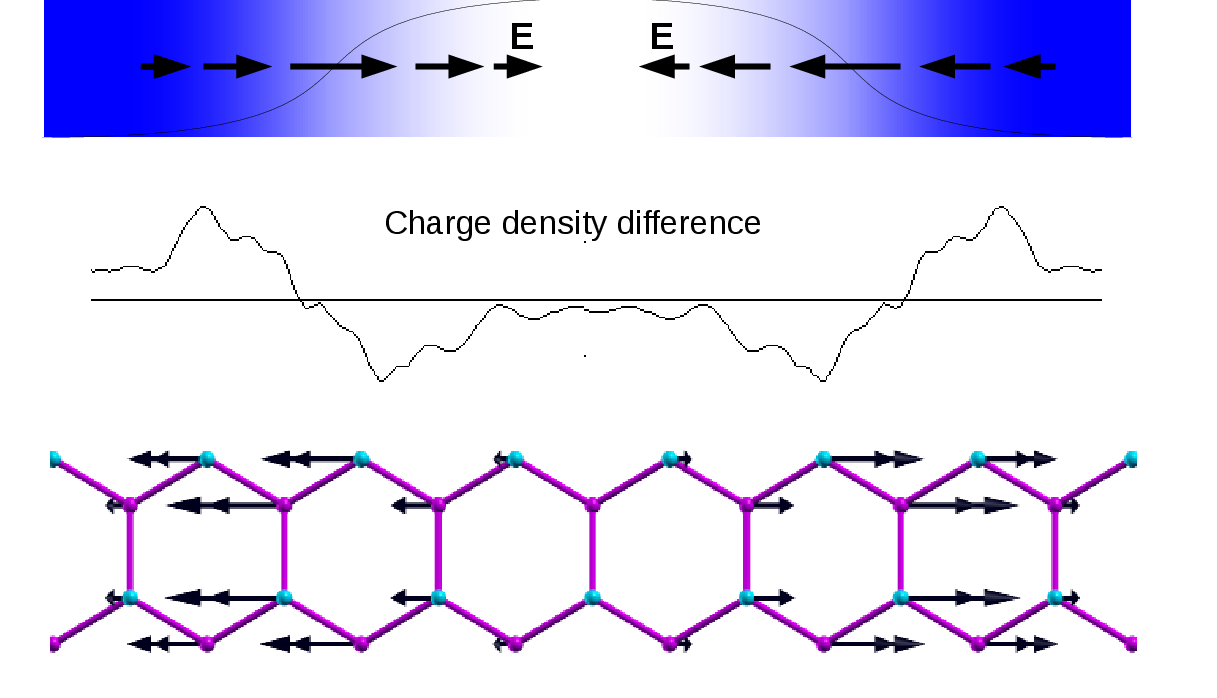}
	\caption{(a) The applied bias potential and two views of the relaxed structure of germanane 
monolayer under the influence of the plotted bias. Unit-cell is marked in black dashed line. 
Bias($V_{bias}$) is kept constant over a length of 4\AA and switched sinusoidally to zero over 
lengths of 7\AA on two sides.
    (b) Schematic description of the electric field and difference between charge densities
with and without the potential applied.
The black arrows in the panel below denote the magnitude of actual displacement with $V_{bias}=1.22$V. }
\label{mechanism}
\end{figure}

\section{Computational details}
%
Primary computation in this work includes - (1)  variable cell relaxation of non-uniformly 
biased supercells of Xane comprising of seven or more unit-cells[Fig.\ref{mechanism}(a)],
(2)  fully relativistic calculation of the ground state, and 
(3)  estimate \ztwo  invariant from evolution of
centres of hermaphrodite Wannier functions(HWF) as function of component of $\vec{k}$ transverse to the
direction of localization of HWFs.

Electronic structure of the ground state of a given configuration has been computed from first principles
using the plane wave based implementation of density functional theory (DFT)
within the generalized gradient approximation of the PBE exchange-correlation functional, 
as available in the Quantum Espresso code \cite{quantumespresso}.
Variable cell structural optimizations have been performed using the BFGS scheme 
until the forces on each atoms is less than 0.001 Ry/au.
%
%
Alternating biased and unbiased regions resembling channels of alternating bias,
as realistically applicable using parallel lines of electrodes, have been considered as shown in Fig.\ref{mechanism}(a). 
%
%
Fully relativistic pseudo-potentials allowing non-collinear calculation have been used to compute band-structure and 
Kohn-Sham spinor wavefunctions. 
Semi-core states have been considered in case of Sn.
In all calculations, k-point grid densities are equivalent to an 18x18x1 k-mesh for the primitive germanane unit-cell. 
%
%
%

For estimation of \ztwo invariant, non-Abelian matrix generalization of the Berry phase is computed for the multitude of occupied bands across the Brillouin zone(BZ) along the two orthogonal directions parallel to reciprocal lattice vectors $\vec{b}_1$ and $\vec{b}_2$ as functions of $k_2$ and $k_1$ respectively where $\{1,2\}\equiv\{x,y\}$. Eigenvalues of the matrices evaluated along  $k_x$($k_y$) for a given $k_y$($k_x$) are centres of the 
HWF with maximum localization in $x$($y$) direction but Bloch like in the $y$($x$) direction.
Following the Thouless topological pumping\cite{fu2006time} of charge, 
\ztwo invariant is estimated\cite{SoluyanovWannierRep, SoluyanovNonInversion} by noting 
the number of intersection of trace of centres of HWFs localized along  $x$($y$) direction as function of 
$k_y$($k_x$) over half the BZ.

For mechanistic understanding of the observed onset or promise of topological phase, 
we compared the nature of evolution of band structures and states 
at the edges of valence and conduction bands at TRIM point with 
those in gapped graphene monolayer and ribbons  at half-filling within the 
Kane-Mele model\cite{TI_QSHE_KaneMele} with inhomogeneity in parameters 
without spin mixing, implying Hamiltonian of form:
\begin{eqnarray}
H&=&\sum_i (m_{\tau_i}+ E_i)c^\dagger_i c_i + t\sum_{\langle ij\rangle}(c^\dagger_i c_j+h.c.)\nonumber\\ 
& &+\lambda\sum_{\langle\langle ij\rangle\rangle}
(ic^\dagger_i c_j+h.c.)
\label{kanemeleeqn}
\end{eqnarray}
for each spin with the second neighbour hopping amplitude $\lambda$ having opposite sign for the two spins.
The mass term $m_{\tau_i}$ is set to +0.7eV  and -0.7eV \cite{Vanderbilt_2018}
for the two sublattices ($\tau$).
The onsite term $E_i$ and hopping amplitudes $t$ and $\lambda$ are varied to represent differently biased regions.
\begin{figure}[b]
\includegraphics[scale=0.32]{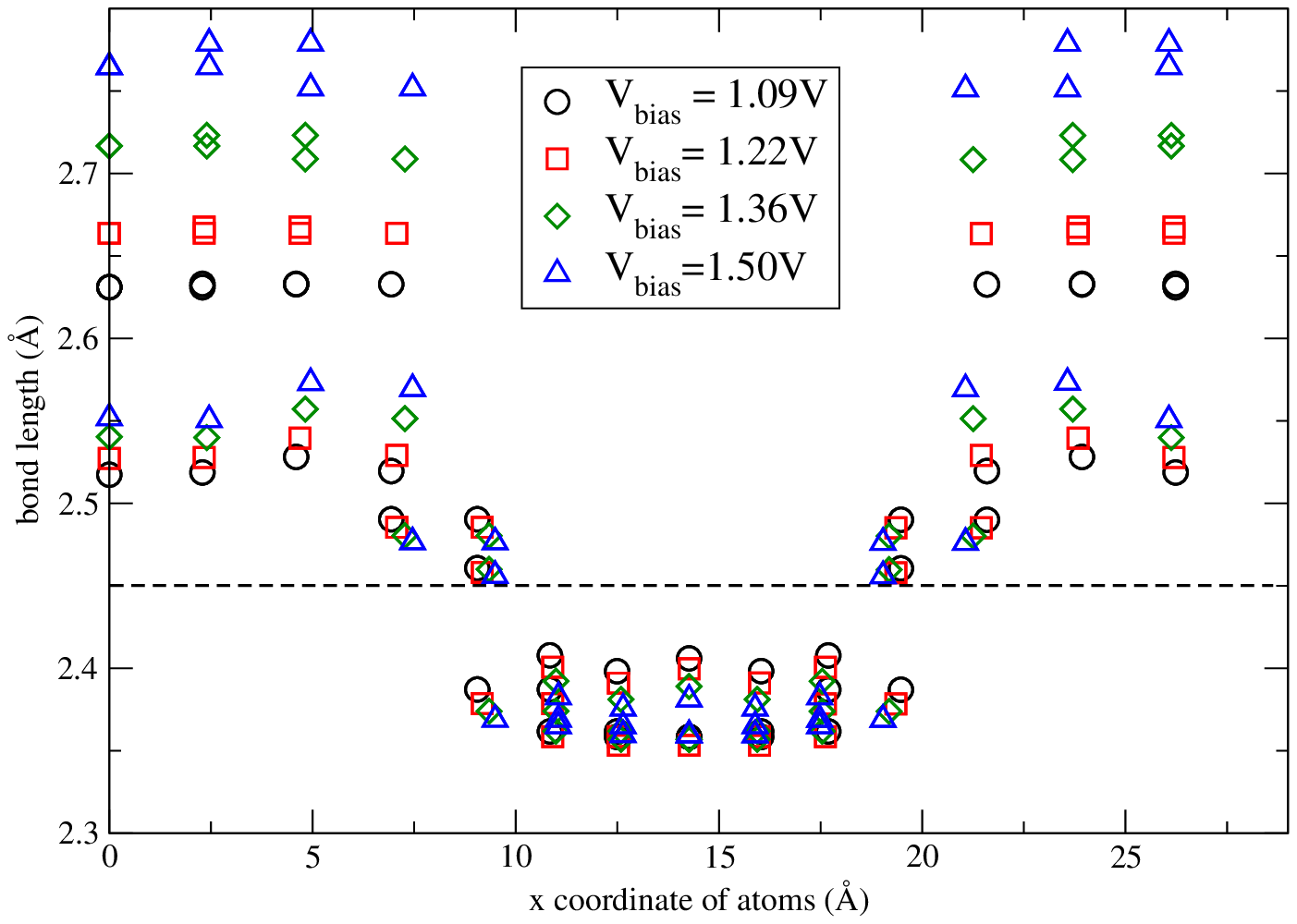}
	\caption{Ge-Ge bond-lengths around each Ge atom across the relaxed germanane super-cell[Fig.\ref{mechanism}(a)] 
for different bias amplitudes($V_{bias}$) with resultant in-plane electric field component in the zigzag direction.
}
\label{bondlengths}
\end{figure}
\begin{figure}[b]
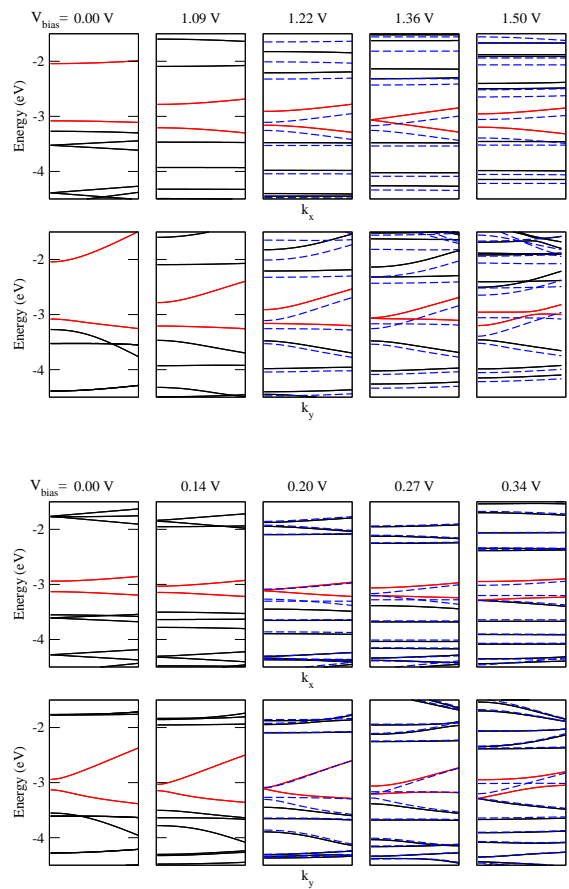

\flushleft (a) Biased germanane:\\ \centering 
	\includegraphics[scale=0.32]{plotbs_GeH_noFermishift.eps}
\flushleft (b) Biased stanane \\ \centering 
	\includegraphics[scale=0.32]{plotbs_SnH_noFermishift.eps}
	\caption{Band-structures of inhomogeneously biased germanane and stanane respectively are plotted in (a) and (b)
for 56 atom orthogonal super-cell[Fig.\ref{mechanism}].
        The range of $k_x$ and $k_y$ plotted cover a fifth of respective BZs from $\Gamma$. The bands shown in black and red
        are computed with SOC applied.   
        Valence and conduction bands are marked in red.
        Bands shown in dashed blue are computed without any relativistic correction.}
\label{bandstructure}
\end{figure}
\begin{figure}[t]
\includegraphics[scale=0.45]{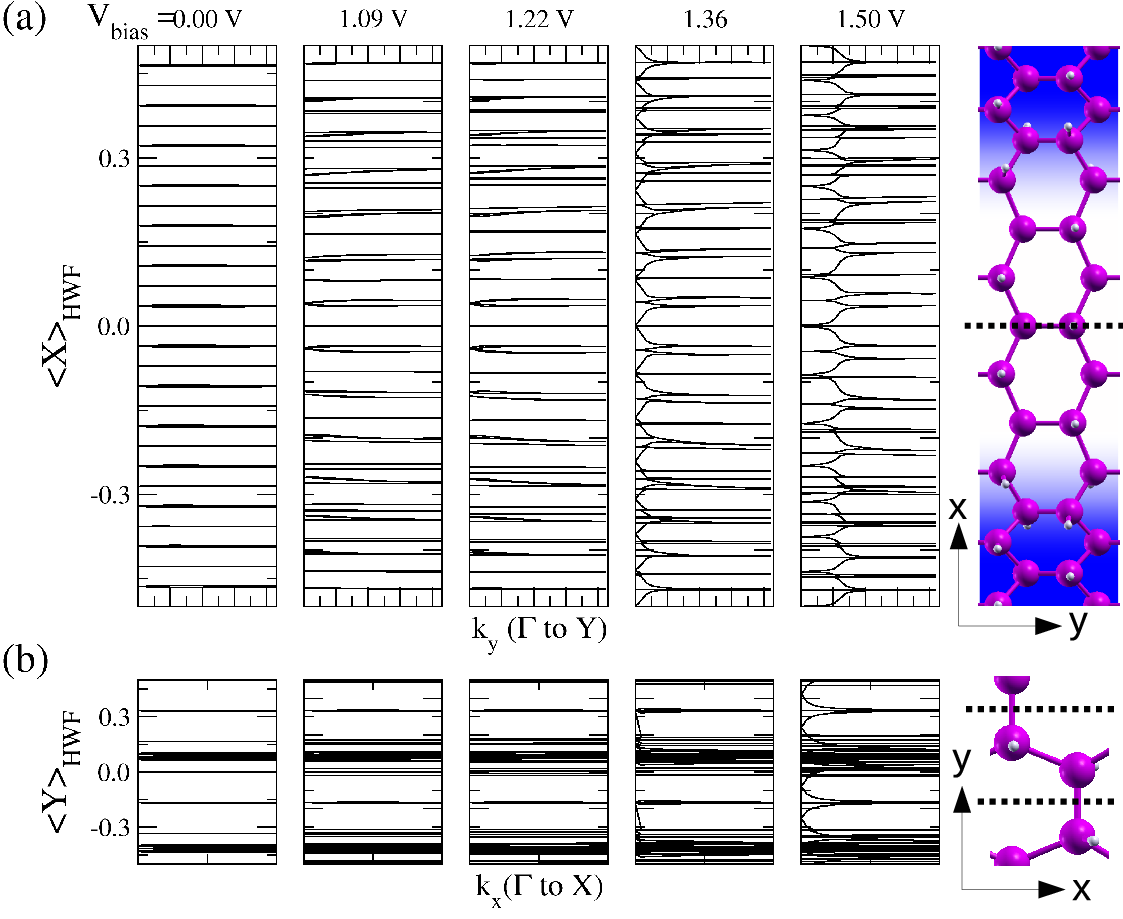}

    \caption{
    (a) Centres of HWFs localized along $\hat{x}$ as a function of $k_y$ in germanane super-cell[Fig.\ref{mechanism}].
    (b) Counterpart of (a) for HWFs localized along $\hat{y}$. 
Location of atoms are provided as guide to eye.
    }
\label{germanane}
\end{figure}
\begin{figure}[t]
\includegraphics[scale=0.72]{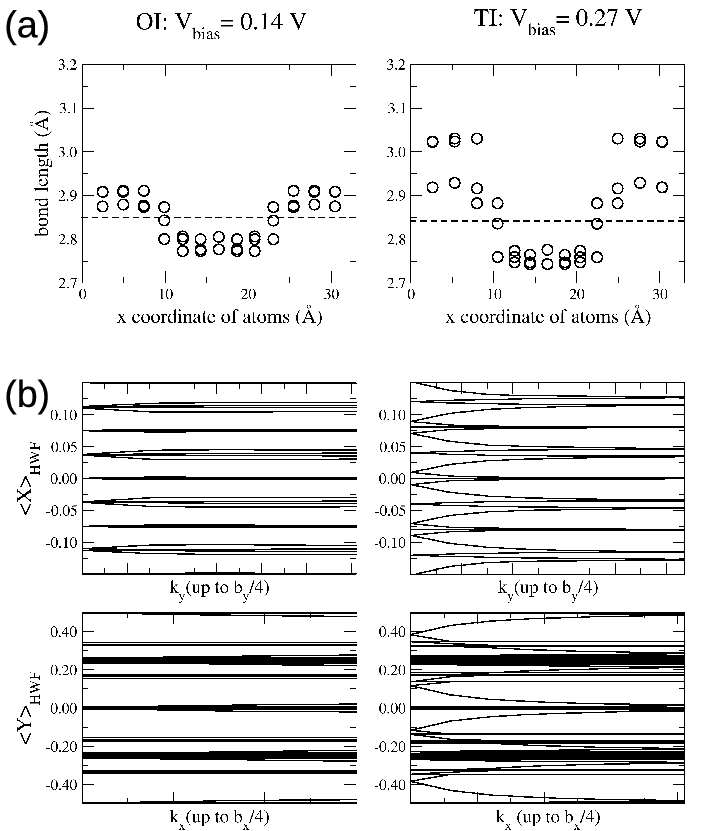}

\caption{(a) Sn-Sn bond-lengths around Sn atoms across the relaxed stanane super-cell similar as in  Fig.\ref{mechanism}(a) before and after transition to TI phase.
    (b) Centres of HWFs localized along $\hat{x}$($\hat{y}$) at upper (lower) panels as function of $k_y$($k_x$) 
spanning over a quarter of the BZ in respective directions.
Plots in the left(right) column corresponds to $V_{bias}=$ 0.14V (0.27V).
}
\label{stanane}
\end{figure}
\begin{figure}[t]

\flushleft (a) \\ \centering \vspace{-0.050 cm}\hspace{0 cm}  
    \includegraphics[scale=0.28]{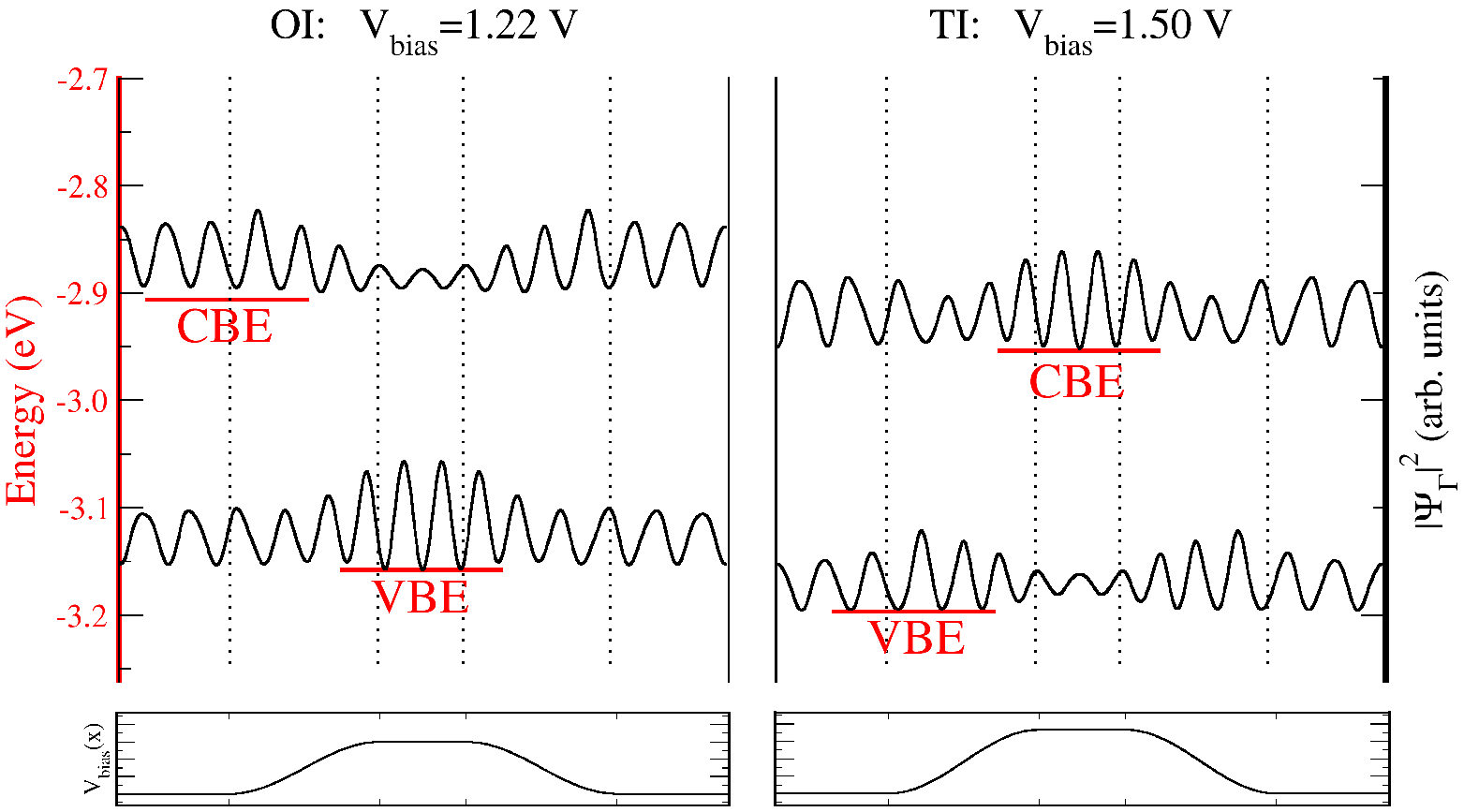}\\  
    \vspace{0.00 cm}\hspace{0 cm}
    
\flushleft (b) \\ \centering \vspace{0.00cm}   \hspace{-4.4cm}  
    \includegraphics[scale=0.28]{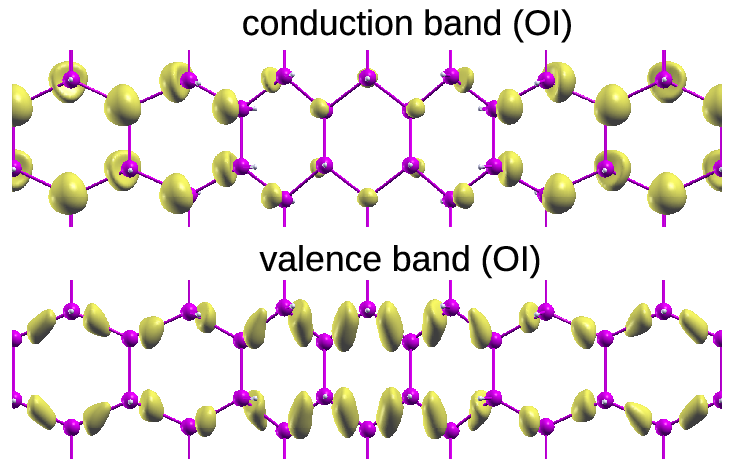}\\
    \vspace{0.0cm}  \hspace{0.0cm} 
    
\flushleft        \centering \vspace{-3.7 cm}   \hspace{0.0cm}  
          (c)\\              \vspace{0.0cm}    \hspace{4.0cm}
    \includegraphics[scale=0.28]{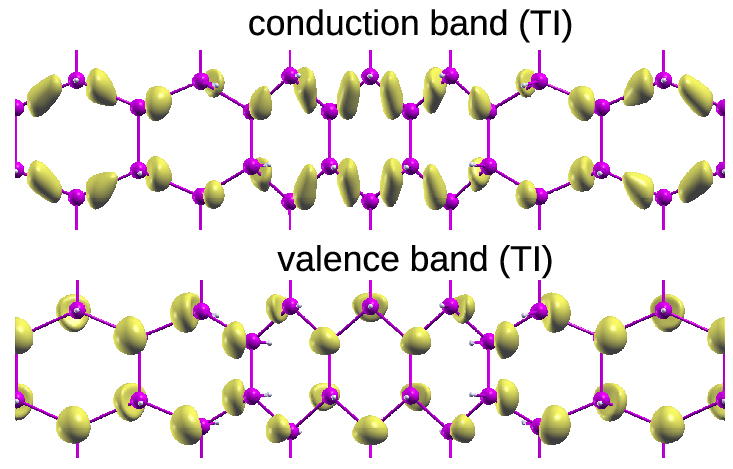}
    \vspace{0.0cm}  \hspace{0.0cm}
    
\flushleft  (d)\\ \centering \vspace{-0.20 cm}  \hspace{0.00cm}
    \includegraphics[scale=0.50]{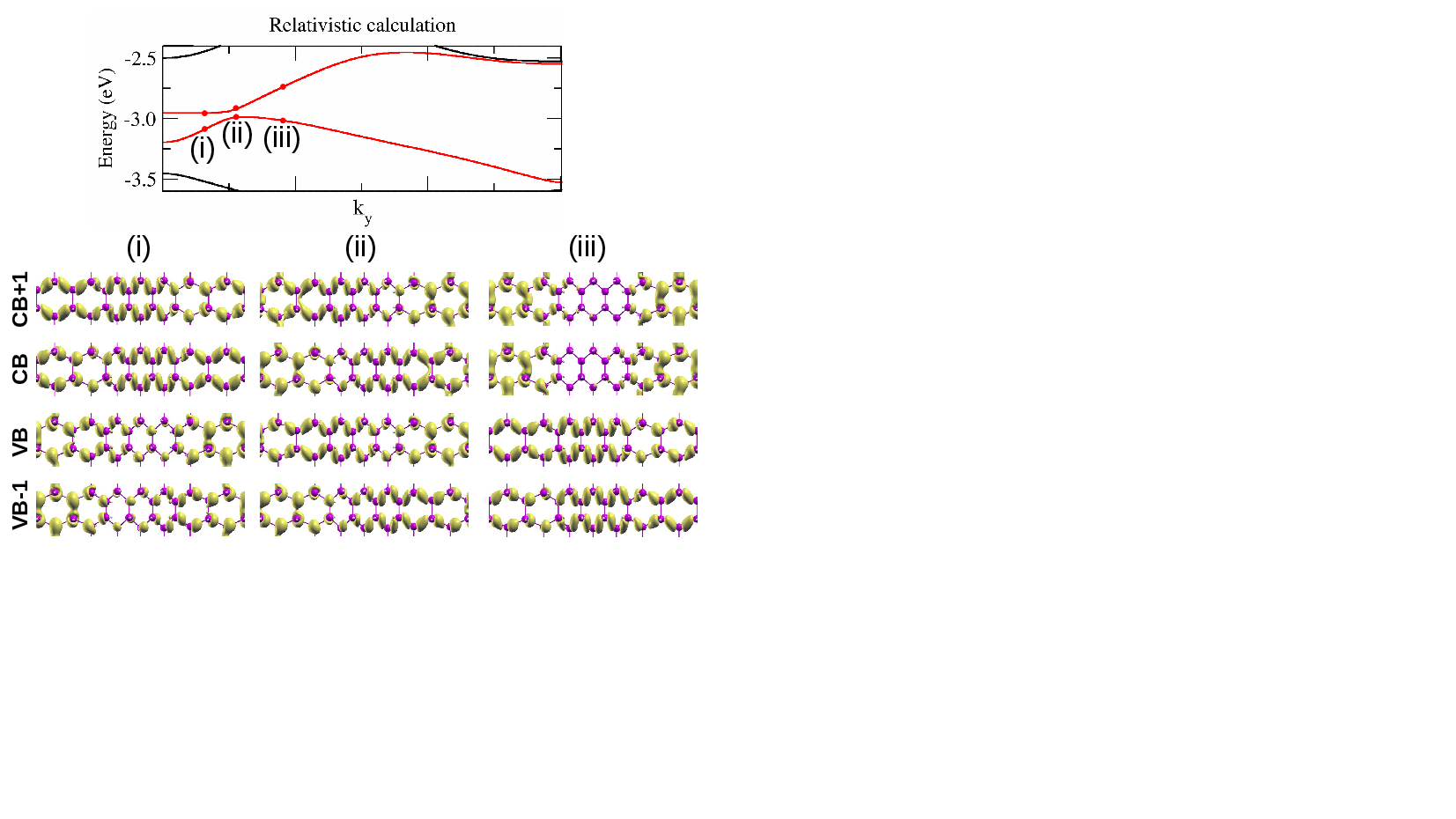}
    \vspace{0.0cm}  \hspace{0.0cm}
    
\flushleft  (e)\\ \centering \vspace{-0.20 cm}  \hspace{-0.00cm}
    \includegraphics[scale=0.50]{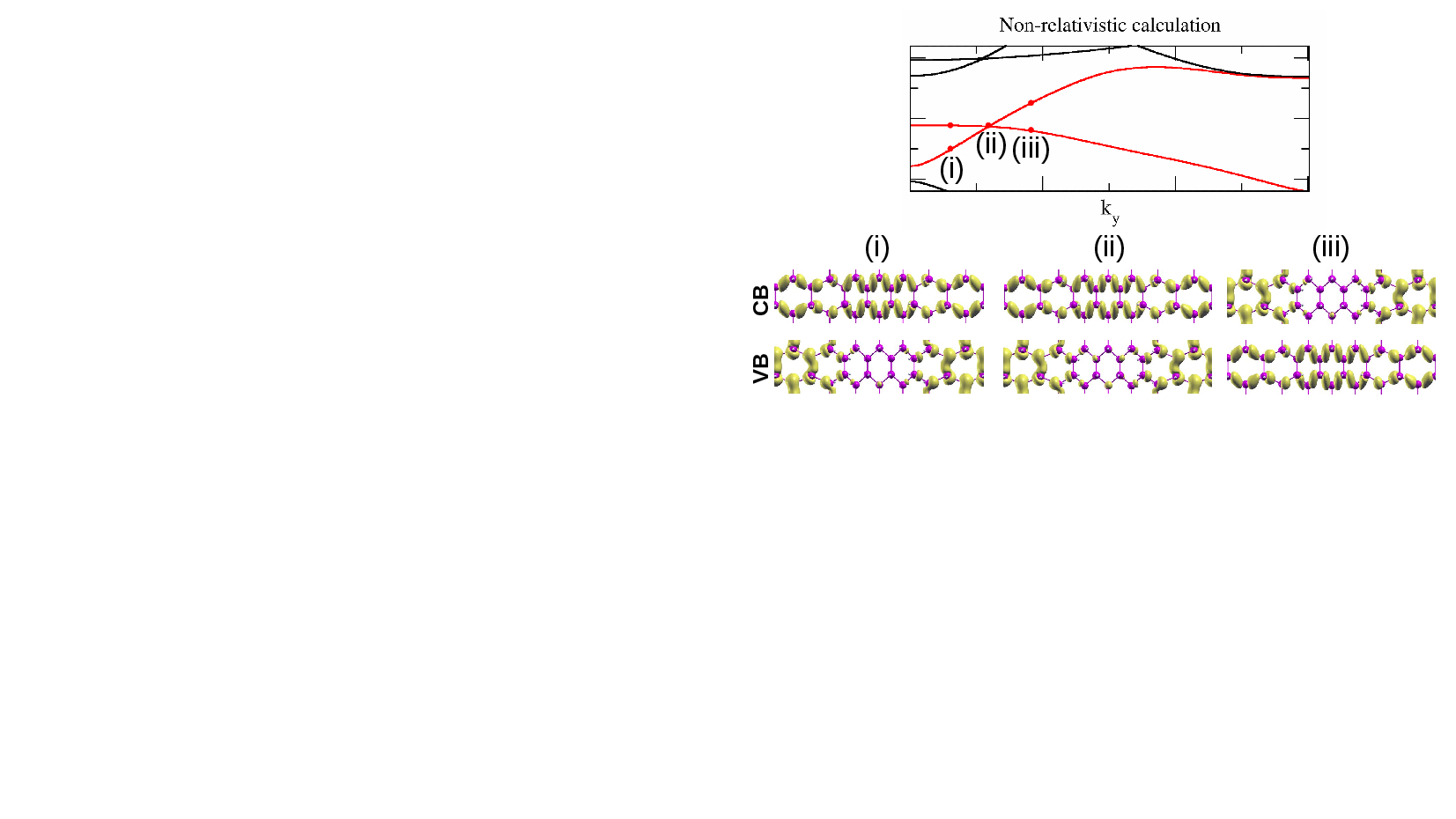}
    \vspace{0.0cm}  \hspace{0.0cm}
    
    \caption{(a) Projection of VBE and CBE at TRIM ($\Gamma$) on $\hat{x}$ direction for germanane super-cell[Fig.\ref{mechanism}(a)] before and after TI transition. VBE and CBE at $\Gamma$ for (b)OI at 1.22V and (c)TI at 1.50V.     
  VBE and CBE plotted at different k-points from (d)fully relativistic and (e)non-relativistic calculations
 for $V_{bias}=1.50$ V. 
}
\label{vbecbe}
\end{figure}

\begin{table*}[t]
\centering
\caption{Strain(\%) in bonds and lattice constants due to applied fixed uniaxial strain.}
\label{tabone}
\begin{tabular}{cc|c|c|c|c}
  & &\multicolumn{2}{c|}{Germanane}  & \multicolumn{2}{c}{Stanane} \\ 
\begin{tabular}{c}
\includegraphics[scale=0.27]{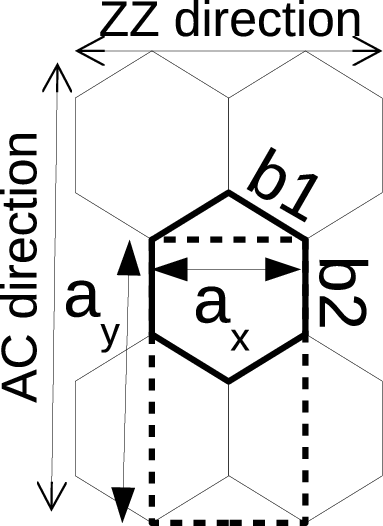}
\end{tabular}
&
\begin{tabular}{c}
strain\\ \hline
in a$_x$ \\ \hline 
in a$_y$   \\ \hline 
in b$_1$ \\ \hline
in b$_2$ \\ \hline
Type \\ \hline
\end{tabular}
&
\begin{tabular}{c|c|c|c}
\hline
\multicolumn{4}{c}{strain applied along ZZ}\\ \hline
 8 & 10 & 12 & 14 \\ \hline 
-1.70 & -2.11 & -2.51 & -2.94  \\ \hline 
 3.56 & 4.48& 5.43& 6.50\\ \hline
0.29 & 0.46& 0.64& 0.66\\ \hline
 OI & OI & TI & TI\\ \hline
\end{tabular}
&
\begin{tabular}{c|c|c|c}
\hline
\multicolumn{4}{c}{strain applied along AC}\\ \hline
-1.26 & -1.58 & -1.76 & -1.92  \\ \hline 
 8 & 10 & 12 & 14 \\ \hline 
 1.48 & 1.98 & 2.36 & 2.72\\ \hline
 4.90 & 5.96 & 7.52 & 9.20\\ \hline
 OI & TI & TI & TI\\ \hline
\end{tabular}
&
\begin{tabular}{c|c|c|c}
\hline
\multicolumn{4}{c}{strain applied along ZZ}\\ \hline
 2 & 3 & 4 & 5 \\ \hline 
 -1.80 & -2.07 & -2.36 & -2.63 \\ \hline 
 0.35  & 0.71  & 1.09  & 1.48 \\ \hline
 -0.45 & -0.39 & -0.34 & -0.29\\ \hline
 OI & OI & OI & TI\\ \hline
\end{tabular}
&
\begin{tabular}{c|c|c|c}
\hline
\multicolumn{4}{c}{strain applied along AC}\\ \hline
 -1.73 & -1.96 & -2.17& -2.39  \\ \hline 
 2 & 3 & 4 & 5 \\ \hline 
 -0.16 & 0.01 & 0.19 & 0.38 \\ \hline
 0.64 & 1.11& 1.61 & 2.12\\ \hline
 OI & OI & TI & TI\\ \hline
\end{tabular}
\end{tabular}
\end{table*}
\section{Results and discussion}
In keeping with our aim to study electro-mechanical actuation, 
we envisage arguably the simplest scenario of in-plane electric field  as evident in Fig.\ref{mechanism},
wherein, parallel linear channels of gates embedded under the Xane layer induce in-plane component of
electric field perpendicular to the gates.
Voltage is smoothly switched from the unbiased region (marked in white) to the 
positively biased region (marked in blue) using a cosine function. 
Notably, bias here onwards is always referred with respect to positive charge.
The resultant electric field varying perpendicularly to the gates would thus cause varying uniaxial strain in 
the same direction, as  evident in the relaxed structure
shown in Fig.\ref{mechanism}(a).
%
%
Unit-cells are thus relaxed parallel to the bias channels in order to allow Poisson's ratio to take effect.

To understand the observed distribution of strain in X-X bonds implied by the relaxed structure we note that 
a covalent bond manifests itself in terms of inter-atomic sharing of electrons. The shared charge directly renders bond-order as introduced by Mayer\cite{Mayer_BO_CPL} implying
a quantification of ``strength'' of a covalent bond. As evident from the charge density difference in Fig.\ref{mechanism}(b), there are two different type of electron transfers - one is grossly bias driven from the regions of  lower to higher bias, which are unbiased and positively biased regions in this work, and the other is occuring locally within  the region of switching of bias under the influence of in-plane electric field present in that region. 
Roughly in one half of the switching region electrons accumulate on the bonds while in the other half the bonds get depleted of charge. The resultant accumulation(depletion) of charge 
between two nearest atoms would in turn imply increase (reduction) in 
the amount of charge shared between atom, leading to increased (reduced) proximity of the two coordinating atoms. This is evident in the degree of displacement of atoms upon relaxation
shown by the black arrows in the  panel at the bottom of Fig.\ref{mechanism}(b).
Thus, narrower the region of switching, higher will be the degree of strain. On the other hand, the variation in bond-order on account of the bias driven gross transfer of electrons results in shorter(longer) bonds in the region of higher(lower) bias,
causing an overall displacement of the atoms in the switching region away from lower bias.  
This also importantly implies that larger the inequality in the width of the regions of lower and higher bias, larger the strain in bonds in the shorter region. Thus a rapidly increasing tensile stain with narrowing of width of the region of lower bias 
paves the way for curated cleavage of bonds.

Based on orientation of X-X bonds w.r.t the direction of electric field
we expect the  b2 bonds (marked in Table.\ref{tabone}) 
to be subjected to maximum uniaxial strain due to electric field applied in the AC
direction.
Whereas with electric field along the ZZ direction, the b2 bonds being perpendicular to the field, 
will not be directly impacted.
Also arguably, strain in b1 bonds(Table.\ref{tabone}) due to  electric field along the ZZ will be less than that 
in the b2 bonds with field along AC.
Thus we expect application of electric field in the ZZ direction to be more favorable
in terms of lower overall strain in the X-X bonds. Indeed we find that the b2 Ge-Ge bonds cleave around a bias amplitude of 0.68 V 
for field in the AC direction, whereas for field in ZZ direction the bonds sustain 
a bias amplitude in excess of 1.5 V. 
Bond length distribution plotted in Fig.\ref{bondlengths} for electric field in ZZ direction shows the degree of
strain to be dominated  expectedly by the b1 bonds in the region of lower bias rising 
to about 12\% before cleavage.
The degree of extension(contraction) of b1 bond length 
increases(decreases) in the region of lower(higher) bias with increasing bias difference.
This slowing down of bond contraction in the higher biased region is because of rapid increase in 
Coulomb repulsion in contracting bonds.
%

\begin{figure}[t]
\flushleft (a)\\ \centering \vspace{-0.25cm}\hspace{-0.0cm}
\includegraphics[scale=0.28]{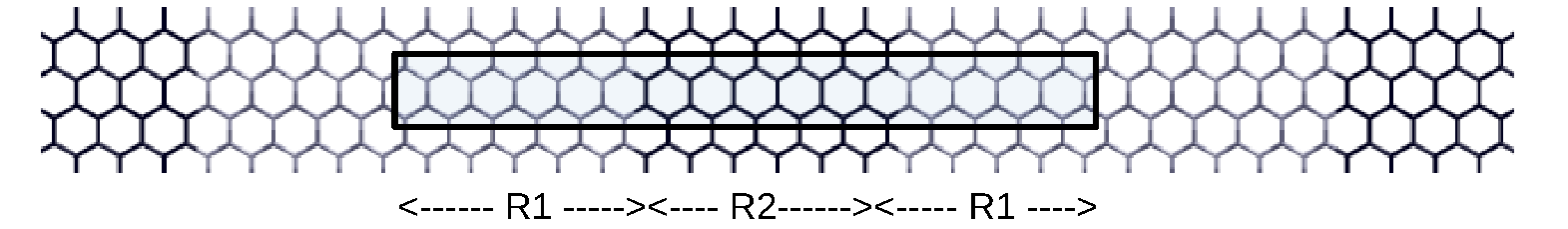}\\
\flushleft (b)\\ \centering \vspace{-0.23cm}\hspace{-0.20cm}
\includegraphics[scale=0.38]{kenemele_bulk.eps}
\caption{(a) Unit-cell of the gapped graphene monolayer model system with two regions R1 and R2 per unit-cell. 
(b) Band-structure and charge density of valence and conduction band edges at $\Gamma$ 
computed with different on-site terms for region R1. }
\label{schema}
\end{figure}

Next we first qualitatively compare[Table\ref{tabone}] the degree of uniform uniaxial strain required to induce 
topological transition in sheets of germanane and stanane within the PBE exchange-correlation.
With  biaxial tensile strain,
germanane is reported to undergo topological transition with 
strain of around 10\%-12\%\cite{GeH_ChenSi,li2014tuning},
whereas stanane is expected to require a strain of about 8\%\cite{liu2016two,lu2017quasiparticle}.
As summarized in Table.\ref{tabone},
onset of TI phase in germanane (stanane) occur between 10\% to 12\% (4\% to 5\%) strain of unit-cell along ZZ which results 
into about 5\% (1.5\%) strain of the two b1 bonds 
and less than 1\% strain of the b2 bonds.
Whereas for strain along AC direction the transition occurs between 8\% to 10\% (3\% to 4\%) strain of the unit-cell leading to 
about 6\% (1.5 \%) strain in the b2 bond 
and 2\% (less that 1\%) strain in the b1 bonds.
Thus, although the degree of uniaxial strain required for onset of TI phase is lower if applied along AC,  
the degree of resultant strain in the individual bonds at the onset of TI are
non-nominally higher than that required if strain is applied along ZZ, more so for germanane than for stanane.
This is also evident from the fact that the contraction of the unit-cell in
the transverse direction is consistently larger in case of ZZ strain implying higher mitigation
of the applied tension. 
Thus from the perspective of overall  strain in the bonds required for the onset of TI phase,
mechanically induced strain along ZZ direction appears to be a more favorable scenario. 
In fact strain along ZZ  induced electrically has been been already discussed above to be favorable 
in terms of sustaining structural integrity.

We therefore here onwards in this work consider electric field in ZZ direction as applicable through 
bias channels in the AC direction (Fig.\ref{mechanism}(a)).
Substantial lowering of band gap(Fig.\ref{bandstructure}(a)) all the way to gap closure and eventual
inversion of bands is observed as a function of increasing magnitude of the bias difference.
As evident from the evolution of centres of HWFs (Fig.\ref{germanane}(a)) in germanane, 
onset of a {\it weak} TI phase occurs with the bias potential 
between 1.22V to 1.36V, marked by \ztwo\ oddness 
only for HWFs localized in $\hat{x}$ direction as a function of $k_y$, 
consistent with the band inversion observed in the same direction (Fig\ref{bandstructure}a).
\ztwo oddness as a function of $k_x$ is observed at 1.50V(Fig.\ref{germanane}(b)) marking onset of a full TI phase. 
%
Bond-lengths plotted in Fig.\ref{bondlengths} for germanane suggests that the emergence of the TI phase occurs after the 
tensile(contractile) strain in the b1 bonds in the lower(higher) biased region exceeds 10\%(4\%), 
amounting to an average tensile strain of about 5\%  which is consistent with
the strain reported in Table.\ref{tabone} to be necessary for emergence of the TI phase 
under homogeneous uniaxial strain along ZZ direction.
Stanane turns into full TI with bias potential 0.27V (Fig.\ref{stanane}) 
marked by \ztwo oddness due to evolution of HWCs in both $\hat{x}$(Fig.\ref{stanane}b) and 
$\hat{y}$(Fig.\ref{stanane}c) directions,  with about 3.5\% average tensile strain
consistent with that  reported in Table.\ref{tabone}.
\begin{figure}[t]
\flushleft \ztwo : (a) as function of $k_x$, \hspace{0.80cm}(b)function of $k_y$ \\ 
\includegraphics[scale=0.20]{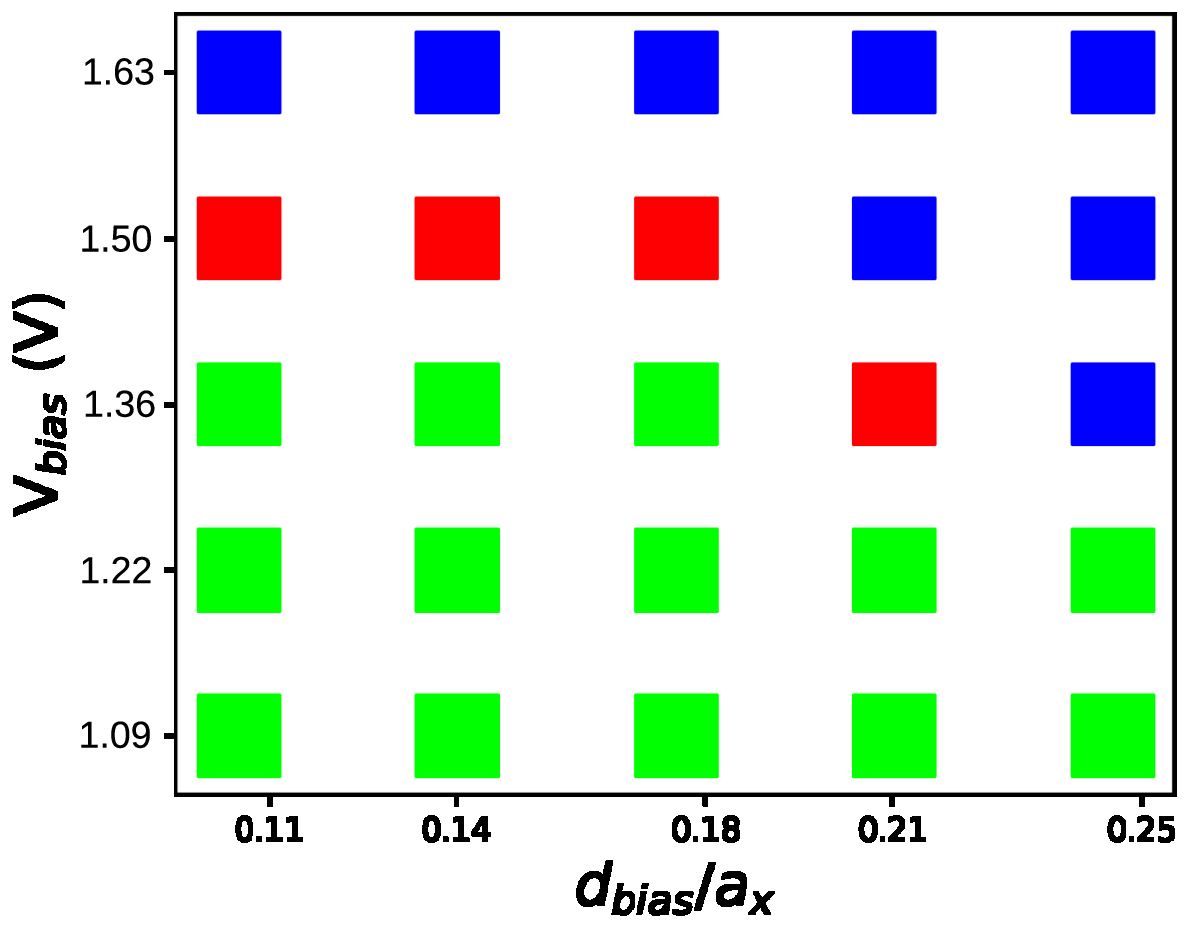}
\includegraphics[scale=0.20]{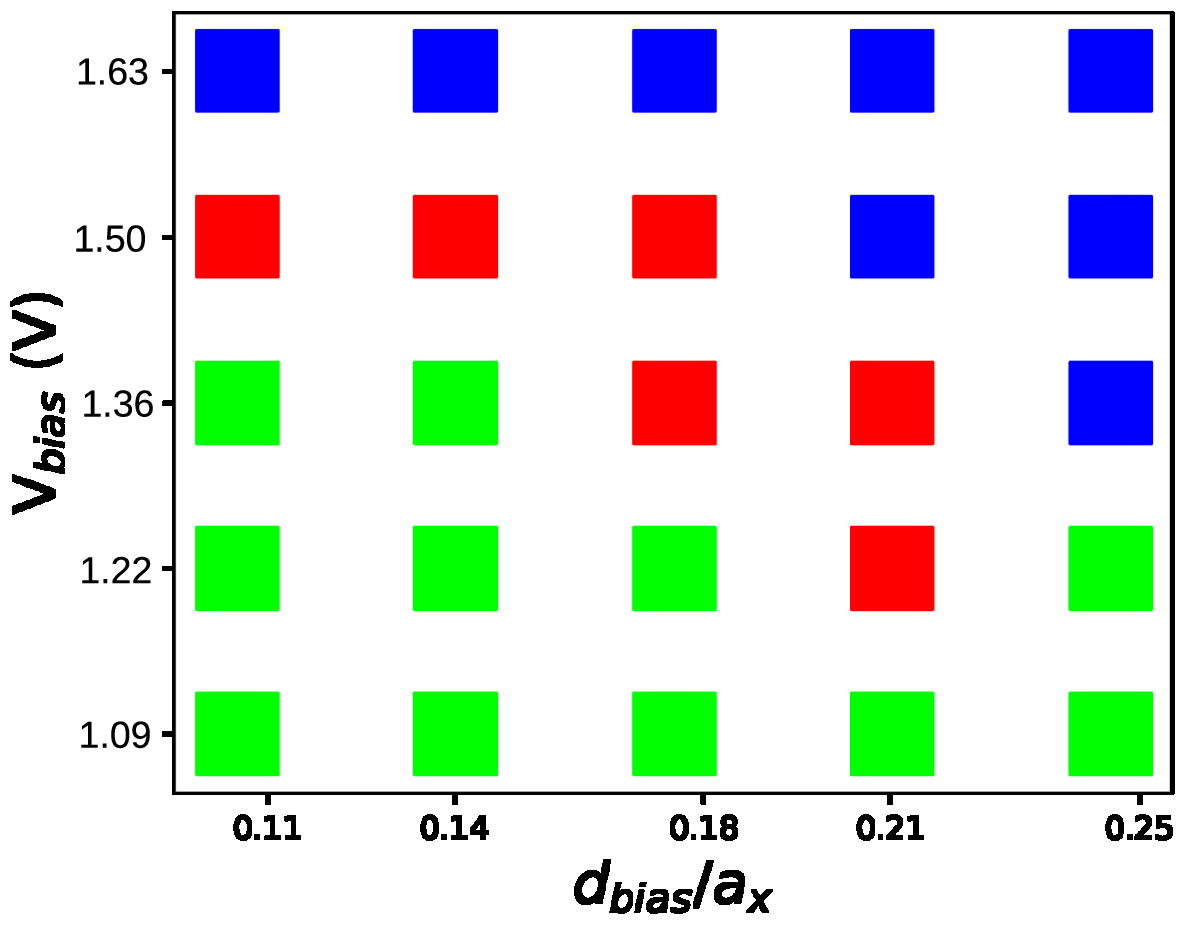}
\caption{Map of emergence of \ztwo\ oddness(marked in red)  
for different bias amplitude($V_{bias}$) and fraction of unit-cell over which bias is applied in 
germanane super-cell[Fig.\ref{mechanism}(a)]. 
Marked in green are scenarios where the system remains trivial. 
In blue marked are scenarios where the system where bonds starts cleaving.}
\label{phasediagram}
\end{figure}
\begin{figure*}[t]
\includegraphics[scale=0.63]{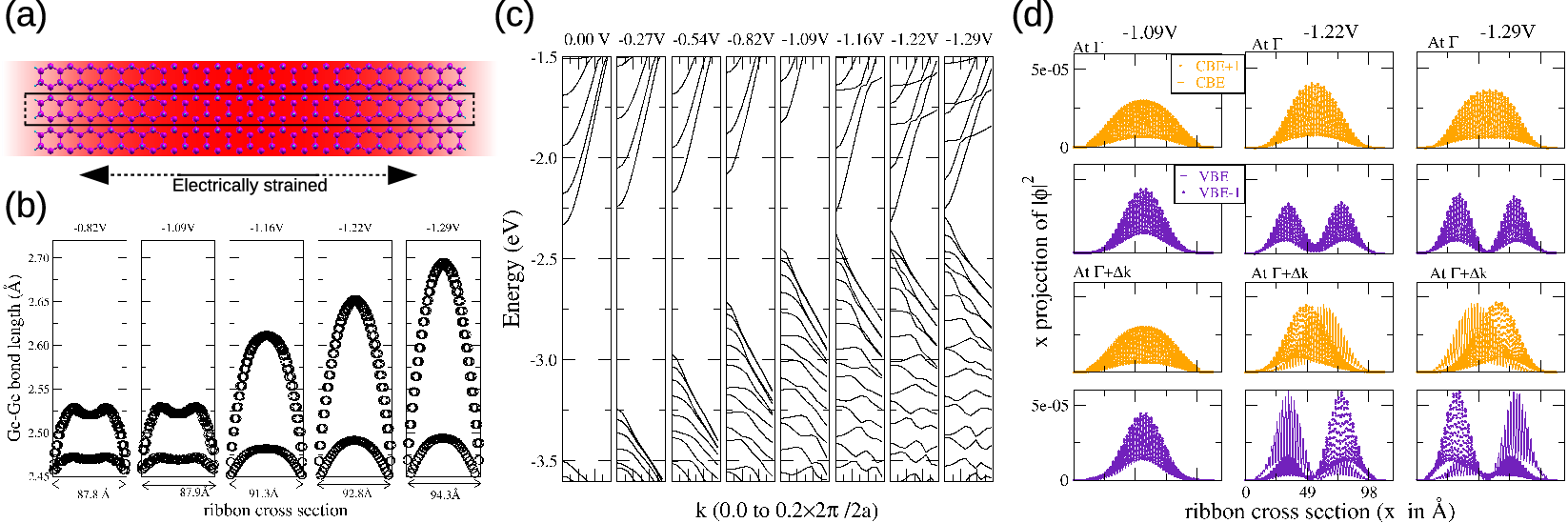}
\caption{(a)In-homogenously biased germanane ribbon of width 20 zigzag units. 
The inhomogeneity of the applied negative bias is depicted as 2D contour in the back-ground. 
(b) Ge-Ge average bondlength plotted along the ribbon cross section. 
(c) Band-structures of inhomogeneously biased germanane ribbon with increasing $V_{bias}$. 
(d) CBE+1,CBE, VBE and VBE-1 at $\Gamma$ and $\Gamma+\Delta k$ plotted for increasing  $V_{bias}$.}
\label{gehribbon}
\end{figure*}
%

Important from the point of detection and application, 
band inversion at TRIM($\Gamma$) 
is also accompanied by spatial swapping of peaks of the valence and conduction band edges(VBE \& CBE) at TRIM.
%
Upon transition to TI phase, substantial population of VBE at $\Gamma$ shifts[Fig.\ref{vbecbe}(a)] from the region of higher bias to the switching region, and vice-versa for the CBE,
while the nature of the VBE and CBE flips at $\Gamma$[Fig.\ref{vbecbe}(b-c)]. 
The nature of the two bands primarily differ in presence and absence of nodes of the
wavefunctions on the path of X-X nearest neighbor coordination, implying expectedly a switch of parity. 
%
%
Interestingly, the observed band inversion and swapping of population of the VBE and CBE at $\Gamma$ 
is also observed 
without any relativistic input like spin-orbit coupling(SOC) or
relativistic correction to kinetic energy, 
implying reasons other than relativistic effects to be responsible for band inversion in the given system. 
This becomes apparent through comparison of VBE and CBE in Fig.\ref{vbecbe}(c) for $\Gamma$
with those in (e)-``(i)" in the same figure depicting VBE and CBE close to $\Gamma$ computed 
without relativistic correction post band inversion at $\Gamma$.  
%
%
As indicated by the band-structure shown in dashed blue in Fig\ref{bandstructure},
with  non-relativistic pseudo-potential, gap-closure happens at $\Gamma$ at a bias close to that with SOC applied, 
followed by a narrow gap in the order of meV or less, opening away from $\Gamma$.
%
%
The inversion in fact can be simply understood as interplay of - 
(1) the enhanced(restrained) lowering of the VBE due to increased(decreased) 
strength of electron hopping along contracted(extended) X-X bonds, and 
(2) the resultant increase(decrease) in on-site energy due to increased(decreased) 
inter-atomic Coulomb repulsion owing to increase(decrease) in proximity of X atoms in the contracted(extended) 
X-X bonds, in the higher(lower) biased region.
Indeed the hopping parameters and on-site terms for the $\sigma$ electrons of the Ge atoms computed in the 
Wannierised basis constructed following templates of hybrid atomic orbitals \cite{JPCA_JBGroup} 
directed along nearest neighbor coordination around Ge atoms show comparable increase ($>$1eV)
in the biased (1.5V) region, compared to their counterparts in the unbiased region. 

Accordingly, to demonstrate the lowering of gap leading to inversion at $\Gamma$ completely due to 
the competing non-relativistic tight-binding parameters rooted at the factors anticipated above, 
we considered a graphene hetero-structure supercell[Fig.\ref{schema}(a)] 
with two regions R1 and R2, of different nn-hopping($t_{nn}$) set to -1.0eV,-1.5eV respectively, 
for a single $p_z$ orbital per site at half-filling.
Notably, electron being negatively charged, the enhancement of on-site term due to reduced X-X separation 
in the positively biased region would be partially cancelled by the bias itself.
However, such cancellation would not happen in the unbiased region where the on-site term reduces due to 
increased X-X separation.
Accordingly, on-site term $E_{R1}$ is reduced over the sites marked as R1 
resembling the unbiased region of the germanane supercell[Fig.\ref{mechanism}(a)]
with stretched Ge-Ge bonds.
Lowering of band-gap and eventual inversion of bands at $\Gamma$ between $E_{R1}=$ 1.6eV and 1.8eV 
is indeed observed[Fig.\ref{schema}(b)] accompanied with spatial swapping of VBE and CBE between R1 and R2 in agreement with that
observed for non-uniformly biased germanane[Fig.\ref{vbecbe}(a-c)].   
%
%
%
%

To examine the role of SOC we note that without SOC the system would have remained an OI 
with a very narrow gap or in-effect a semi-metal.
With fully relativistic pseudo-potential enabling SOC, 
the gap increases to the order of 0.03eV after closure at $\Gamma$[Fig.\ref{vbecbe}(d-e)].
%
Without SOC the nature of the states
[Fig.\ref{vbecbe}(e)] changes abruptly across the narrow gap implying in effect a band crossing with 
only nominal mixing of the bands at the point of crossing. 
%
%
With SOC however we see substantial and systematic mixing[Fig.\ref{vbecbe}(d)] in the nature of 
wavefunctions across the unitcell. 
In particular the states at the gap marked as ``(ii)'' in Fig.\ref{vbecbe}(d) suggests
the degenerate pairs constituting VBEs (and similarly the CBEs) to be  equal but out of phase 
 mixtures of the two parities charaterizing the VBE and CBE at $\Gamma$.
Thus, although SOC in this system is not responsible for band inversion at TRIM,
it is responsible for mixing of bands of different parities and opening of gap 
leading to onset of the topological phase after 
band closure at $\Gamma$.
Thus it is apparent that if band inversion occurs non-relativistically 
within the range of bias potential sustained by the system, 
then a TI phase is likely to exist if SOC consilidates sufficiently in the process to open a gap.
The consolidation would occur primarily at the region of lower bias where stretched
X-X bonds would lower the effective hopping of electron between $\sigma$ orbitals 
and bring the corresponding bonding and anti-bonding manifolds sufficiently close enough
energetically for the SOC of  $\sigma$ orbitals to take effect.
Notably, the presence of electric field in part of the unit-cell 
breaks the time reversal symmetry(TRS), implying that the  TI phase is 
a TRS broken quantum spin Hall(QSH) insulator.

We next survey the variability of the onset bias in terms of the widths of the differently biased regions.
As already discussed, in order to maximize tensile strain in the lower biased region it needs to be narrower. 
On the other hand, wider the higher biased region lower the compressive strain and thus lower the 
proximity of X atoms 
crucial for band inversion.
Thus an optimal ratio of widths of the higher and lower biased regions is essential
for the onset of TI phase at the lowest bias potential given the length of the super-cell,
in order to ensure sustainability of the system.
Accordingly we consider super-cell shown in Fig.\ref{mechanism}(a) and
increase the width of the higher biased region 
($d_{bias}$ in Fig.\ref{phasediagram}) keeping the super-cell length unchanged and track the onset of TI phase.
The width of the switching region is kept at 7\AA.
The variation of onset of TI phase as function of biased fraction of the super-cell and voltage applied is summarised
in Fig.\ref{phasediagram}. 
As expected, with increasing width of the biased region  
\ztwo oddness (marked by red) emerges at a lower positive bias
for both directions - $k_x$ and $k_y$, but up to a critical fraction of about a fifth of the unit cell.
Scenarios marked in blue are the ones where bonds start cleaving on account of narrowing lower biased region.
With larger super-cells TI phase emerges at a lower voltage 
for same biased fraction.
Thus with realistically larger super-cells, implying larger separation and width of the parallel gates, 
the TI phase may be accessible at lower bias potentials well within the limit of sustainability of the monolayers.
\begin{figure}[t]
   \includegraphics[scale=0.32]{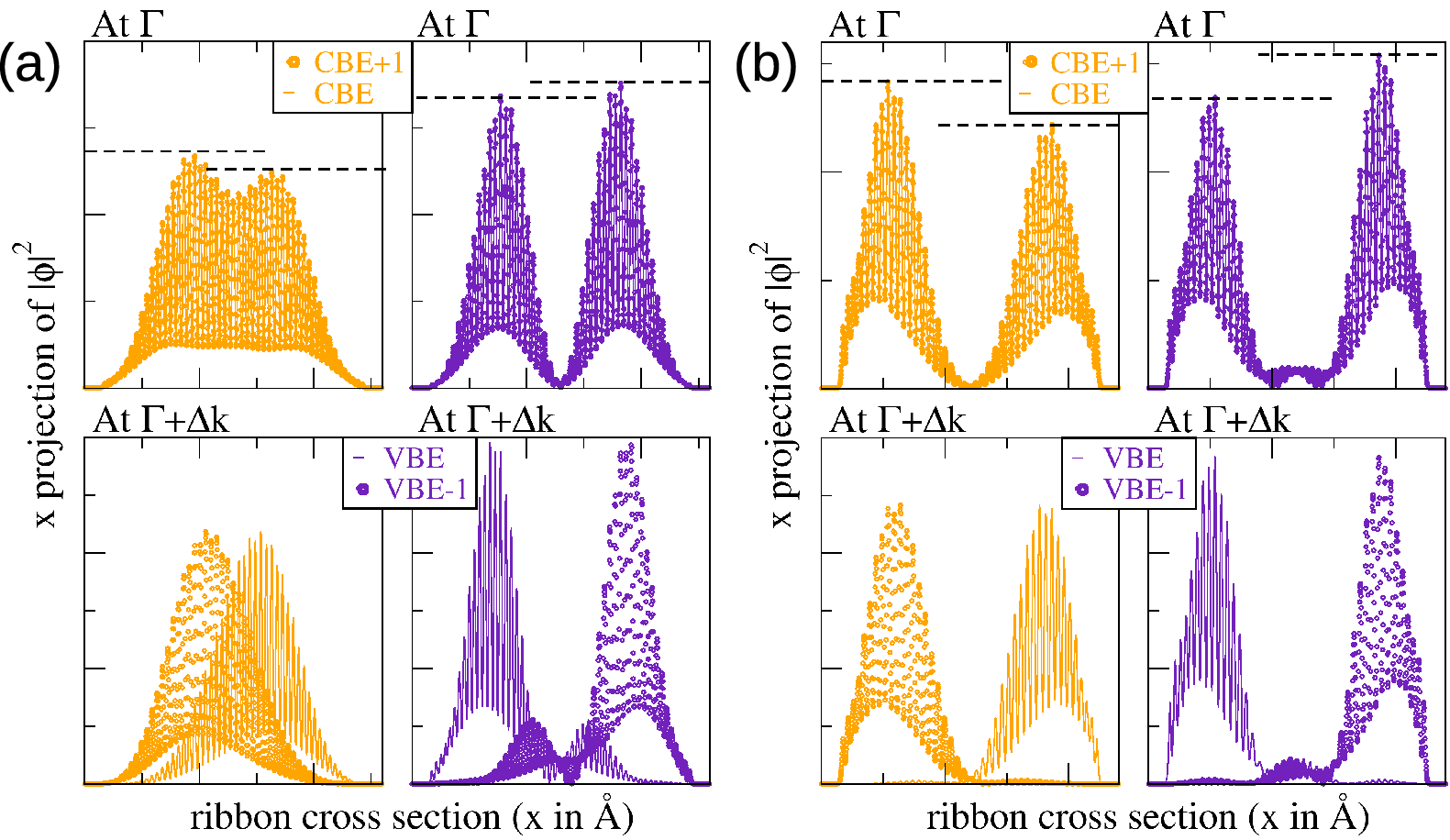}
\caption{ Projection of CBE+1, CBE, VBE, VBE-1 at $\Gamma$ and $\Gamma+\Delta$k plotted in 
(a) and (b) respectively for $V_{bias}=$-1.50V and -0.34V applied halfway along AC edged germanane and stanane ribbons 
of width more than 100\AA(24 zigzag units) and less than 100\AA(20 zigzag units) respectively.}
\label{snhribbon}
\end{figure}
\begin{figure}[t]
\flushleft (a)\\\vspace{-0.5cm}\hspace{0.5cm}
    \includegraphics[scale=0.38]{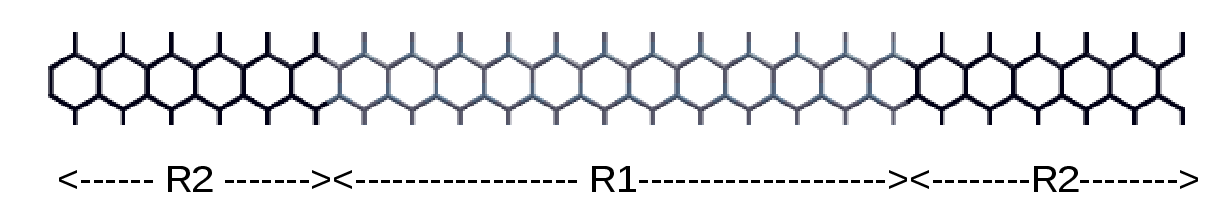}\\
\flushleft (b)	\includegraphics[scale=0.35]{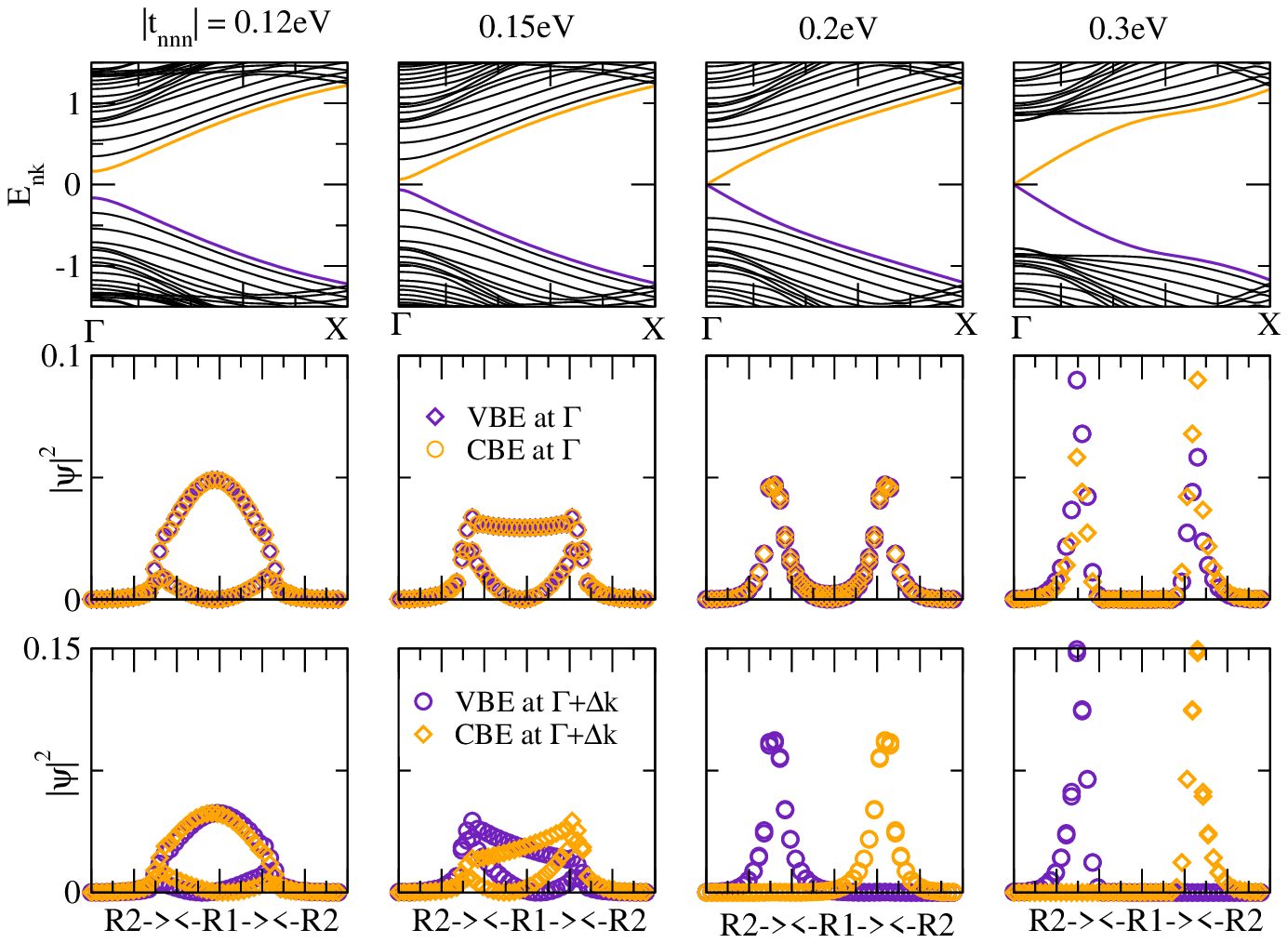}
\caption{(a) Unit-cell of the gapped graphene ribbon model system with two regions R1 and R2. 
(b) Band-structure and charge density of the valence and conduction band eges at $\Gamma$ and a 
neighboring point $\Gamma+\Delta k$.}
 \label{schemaribbon}
\end{figure}

In case of ribbons, we have two options to place the gates - perpendicular or parallel to the ribbons.
Gates perpendicular to ribbons would induce tensile strain along the length of the ribbons, 
which will be substantially compensated by shrinkage of width due to free edges, leading to lesser bond strains.
Additionally, ribbon edges at regions of higher bias would substantially buckle to cause structural degeneration. 
We thus realistically only consider gates parallel to ribbons.
%
%
With the electric field preferred in ZZ direction for optimal distribution of strain,  
the choice for ribbon with gate parallel to it must be armchair edged.
If we consider linear gate embedded beneath the ribbons along their length mid way across their width 
then we need to apply negative bias [Fig.\ref{gehribbon}(a)] so that the
X-X bonds transverse to the ribbon direction can undergo tensile strain needed for TI phase. 
We have considered a Gaussian switching of potential to zero from the peak in the middle across the width of the ribbon. 
%
%
Low shrinkage of unit-cell along the length of the ribbon indicates substantial retention of tensile strain of X-X bonds transversely to the ribbons.
However, strain can not be induced uniformly across the width due to tension released by the free edges,
as evident from the bond-length profile in Fig.\ref{gehribbon}(b).
A tension can of course be imagined to be maintained by pulling the ribbon outward transversely to the ribbon
 through additional parallel positively biased gates on two sides.
%
%
However, in this work we limit to considering effect of only one negatively biased linear gate parallel to the ribbon 
embedded halfway across the width Fig.\ref{gehribbon}(a). 
As evident in Fig.\ref{gehribbon}(c), increasing negative bias causes substantial lowering of 
band-gap driven by the electrically induced tensile strain similarly as happens in monolayer.
Bondlength distribution Fig.\ref{gehribbon}(b) shows  that with increasing bias strength the 
maximum tensile strain occurs halfway across the width, paving the way for the possibility of TI phase 
to emerge over a fraction of a width of the ribbon symmetrically about the gate. 
%
%
%
The unit-cell would thus be partitioned into TI and OI regions
implying appearance of chiral interface states 
localizing intermediately between the halfway and edge on two sides.
%
Within the range of width of ribbons we could compute using the resources at our disposal,
we present a promise that for sufficiently larger widths and appropriate selection of bias it will be
possible to induce topologically protected  chiral interface states at $\Gamma$.
For the ribbon depicted in Fig.\ref{gehribbon}, we indeed see localization of VBE but only an onset of asymmetry 
of CBE within the bias sustained by the ribbon. 
This is expected since the highest bond-length occuring in the middle of the ribbon did not exceed 2.7\AA which
appears to be a threshold bond-length for SOC to consolidate sufficiently as per emergence of TI phase in germanane monolayer.
Promise is borne by the increasing difference of peak heights [Fig.\ref{snhribbon}] of VBE and CBE
localizing at the two interfaces. 
For germanane ribbon appreciable difference is observed in ribbons of width above 100\AA [Fig.\ref{snhribbon}(a)].
Expectedly for stanane ribbon similar population difference is observed at smaller width less that 100\AA
[Fig.\ref{snhribbon}(b)].
The observed  spatial separation of degenerate VBE,VBE-1 and similarly of CBE,CBE+1 
around $\Gamma$ (marked as $\Gamma + \Delta k$ in Fig.\ref{snhribbon}),
indicates the onset of consolidation of TI phase within the region of lower bias.  
These results suggests the possibility to reversibly induce a linear channel of TI region
with chiral interface states  within  realizably large monolayer segments
exclusively through application of negative bias minimally at a single linear gate
embedded underneath.

To substantiate the possibility we again resort to the model gapped-graphene ribbon hetero-structure system [Fig.\ref{schemaribbon}]
made of two regions akin to that considered for the monolayer.
To associate TI nature to the region in the middle (R1) within the Kane-Mele model[Eqn.\ref{kanemeleeqn}], 
we increase the magnitude of $|t_{nnn}|$ around its analytic threshold for onset of TI phase which is 
of $|t_0|=m/(3\sqrt{3})=$0.1347eV, where $m$ is the mass term set to 0.7eV \cite{Vanderbilt_2018}. 
The electronic structure in the model is completely sectorized in the two spins since we have not included any 
Rashba term in our model.
As evident in Fig.\ref{schemaribbon}, with increasing $|t_{nnn}|$ beyond $|t_0|$, VBE and CBE at $\Gamma$ started
localizing at the interface of the two regions.
However, chiral separation of the VBE and CBE to the two interfaces at $\Gamma$  starts at a much higher $|t_{nnn}|$ which is
expected to reduce with increasing width of R1.
In agreement with the observed  spatial separation of the degenerate VBEs and CBEs at $\Gamma+\Delta k$ 
in partially biased germanane and stanane ribbons [lower panels of Fig.\ref{gehribbon}(d) and \ref{snhribbon}], 
in this model also, VBEs and CBEs of the two spin sectors at $\Gamma+\Delta k$ 
[panel in the bottom of Fig.\ref{schemaribbon}] and 
their Kramer pairs at  $\Gamma-\Delta k$  starts localizing separately at the two interfaces 
immediately once $|t_{nnn}|$ exceeds $|t_0|$.
At a much higher value of $|t_{nnn}|$ in excess of about 0.25eV we start observing chiral separation of VBE and CBE,
implying equivalence of the paradigm of partially biased Xane ribbons promised above with TI channel embedded within.
In fact, the location of the interface within the half-width of the ribbon can be custom defined by adjusting the amplitude
of the applied bias. 
\section{Conclusion}
We have computationally demonstrated from first principles an exclusively electrical means to reversibly strain 
and tune  band-gap down to closure followed by emergence of topologically insulating(TI) phase in
monolayers and ribbons of heavier hydrogenated Xenes(Xane), 
namely germanane and stanane, 
through application of inhomogeneous bias using gates of realizable length scale.
The key underlying mechanism is band inversion through interplay of the energetics 
inter-atomic tunneling and Coulomb repulsion at the variable strain in X-X bonds in 
regions of different bias, 
followed by opening of band-gap due to consolidation of spin-orbit coupling due to tensile strain in the 
region of lower bias.
Band inversion is also accompanied by a spatial swaping of conduction and valence band in the region
of higher bias at $\Gamma$, which may be relevant to precesion switching applications.
In sufficiently wide ribbons of these Xanes with negatively biased linear gates embedded midway parallel to them, 
a TI region may emerge over a fraction of width with topologically protected chiral interface states shy of the edge.
However, Karmer degernerate pairs of states from the valence and conduction bands about $\Gamma$ starts
localizing at the two interface separately at lower voltages and narrower ribbons, and thus should be easily accessible. 
%
The demonstrated possibility of electrically induced topological insulator phase should be realizable 
in the broad class of two dimensional materials with $p$ as frontier orbitals.
\section{Acknowledgments}
Computations have been performed using facilities funded by the 
Dept. of Atomic Energy(DAE) of the Govt. of India.
JB acknowledges generous funding support from grant No.RIN4001 from DAE, GOI.

\bibliographystyle{unsrt}
\bibliography{bibliography.bib}

\end{document}